\def\gL{\gamma _{_{\ell}}}

\def\V{{ \langle \widetilde{V} \rangle}}
\def\tV{{\widetilde{V}_{\alpha}}}
\def\S{S^{\widetilde{V}}_{\alpha}}
\def\cV{{\widetilde{V}_{\text{calc}}}}

\def\gL{{\gamma_{_L}}}
\def\gG{{\gamma_{_G}}}

\def\bC{{\beta_{_{C}}}}

\def\lJ{{\lambda_{_{J}}}}
\def\nL{{\mathcal{L}}}
\def\nW{{\mathcal{W}}}

\documentclass[aps,prappl,twocolumn,groupedaddress,showkeys,showpacs,superscriptaddress,floatfix,longbibliography]{revtex4-1}
\usepackage{epsfig}
\usepackage{multirow}
\usepackage{amsmath, amssymb,mathtools}
\usepackage{color}
\usepackage{graphicx}
\usepackage{dcolumn}
\usepackage{bm}
\usepackage{subfigure}
\usepackage[utf8]{inputenc}
\usepackage[T1]{fontenc}
\usepackage{tikz}
\usepackage{soul}

\graphicspath{{Figures/}}

\begin{document}

\title{Voltage drop across Josephson junctions for L\'evy noise detection}

\author{Claudio Guarcello\thanks{e-mail: cguarcello@unisa.it}}
\affiliation{Dipartimento di Fisica ``E.R. Caianiello'', Universit\`a di Salerno, Via Giovanni Paolo II, 132, I-84084 Fisciano (SA), Italy}
\affiliation{INFN, Sezione di Napoli Gruppo Collegato di Salerno, Complesso Universitario di Monte S. Angelo, I-80126 Napoli, Italy}
\author{Giovanni Filatrella\thanks{e-mail: filatrella@unisannio.it}}
\affiliation{Dep. of Sciences and Technologies and Salerno unit of CNISM, University of Sannio, Via Port’Arsa 11, Benevento I-82100, Italy}
\affiliation{INFN, Sezione di Napoli Gruppo Collegato di Salerno, Complesso Universitario di Monte S. Angelo, I-80126 Napoli, Italy}
\author{Bernardo Spagnolo\thanks{e-mail:}}
\affiliation{Dipartimento di Fisica e Chimica ``Emilio Segrè'', Group of Interdisciplinary Theoretical Physics, Università di Palermo and CNISM, Unità di Palermo, Viale delle Scienze, Edificio 18, 90128 Palermo, Italy}
\affiliation{Radiophysics Dept., Lobachevsky State University, 23 Gagarin Ave., 603950 Nizhniy Novgorod, Russia}
\affiliation{Istituto Nazionale di Fisica Nucleare, Sezione di Catania, Via S. Sofia 64, I-95123 Catania, Italy}
\author{Vincenzo Pierro\thanks{e-mail: }}
\affiliation{Dept. of Engineering, University of Sannio, Corso Garibaldi 107, I-82100 Benevento, Italy}
\affiliation{INFN, Sezione di Napoli Gruppo Collegato di Salerno, Complesso Universitario di Monte S. Angelo, I-80126 Napoli, Italy}
\author{Davide Valenti\thanks{e-mail: }}
\affiliation{Dipartimento di Fisica e Chimica ``Emilio Segrè'', Group of Interdisciplinary Theoretical Physics, Università di Palermo and CNISM, Unità di Palermo, Viale delle Scienze, Edificio 18, 90128 Palermo, Italy}
\affiliation{CNR-IRIB, Consiglio Nazionale delle Ricerche - Istituto per la Ricerca e l’Innovazione Biomedica, Via Ugo La Malfa 153, I-90146 Palermo, Italy}


\begin{abstract}

We propose to characterize L\'evy-distributed stochastic fluctuations through the measurement of the average voltage drop across a current-biased Josephson junction.
We show that the noise induced switching process in the Josephson washboard potential can be exploited to reveal and characterize L\'evy fluctuations, also if embedded 
in a thermal noisy background. The measurement of the average voltage drop as a function of the noise intensity allows to infer the value of the stability index that characterizes 
L\'evy-distributed fluctuations. An analytical estimate of the average velocity in the case of a L\'evy-driven escape process from a metastable state well agrees with the numerical 
calculation of the average voltage drop across the junction. The best performances are reached at small bias currents and low temperatures, \emph{i.e.}, when both thermally 
activated and quantum tunneling switching processes can be neglected. The effects discussed in this work pave the way toward an effective and 
reliable method to characterize L\'evy components eventually present in an unknown noisy signal.
\end{abstract}

\maketitle

\section{Introduction}
\label{Sec00}\vskip-0.2cm

In the past two decades, the seminal cue of Refs.~\cite{Pek04,Ank05,Tob04} has prompted several experimental setups of noise detectors based on Josephson devices~\cite{Pek05,Ank07,Suk07,Tim07,Pel07,Hua07,Gra08,Nov09,LeM09,Urb09,Fil10,Add12,Oel13,Add13,Sol15,Sai16}. 
More generally, Josephson devices are nowadays often employed for sensing and detection applications~\cite{Pea98,Vou10,Ber13,Ull15,Wal17,GuaBra19,GuaBraSol19}. 
Indeed, a Josephson junction (JJ) is a natural threshold detector for current fluctuations, being essentially a metastable system working on an activation mechanism~\cite{Bar82,Lik86}. 
In a common set-up, the bias current is linearly ramped until the JJ switches to the finite voltage state. When the voltage appears, one measures the current, or the time, 
at which the passage to the resistive state has occurred. Alternatively, the JJ can be biased to a fixed current, and the time it takes to leave the superconducting state is measured. The two methods have advantages and drawbacks, \emph{e.g.}, see Ref.~\cite{Pountougnigni20}. When a JJ is used for noise detection, to mention just a few 
examples along this line, the interesting information content is obtained from the highest moments of electrical noise to investigate the Poissonian nature of current fluctuations~\cite{Pek04}. 
This work explores also the Poissonian charge injection through the study of the third-order moment of electric noise, while Ref.~\cite{Ank05} proposes a study of the fourth-order 
moment of the noise. In Ref.~\cite{Tob04} a Josephson array has been used to estimate the full counting statistics through the analysis of rare jumps induced by current fluctuations. 
Finally, in Refs.~\cite{Lin04,Hei04} the non-Gaussian nature of an external noise is investigated through the sensitivity of the conductance of a junction in the Coulomb blockade regime. 
However, the discrepancies possibly observed with respect to a typical Gaussian response are rather small and an experimental measurement of higher 
moments, beyond the variance, is indeed demanding.

Alternatively, the characterization of non-Gaussian fluctuations can be addressed by analyzing the switching currents distributions~\cite{Gua17,Gua19}. In particular, 
a specific kind of non-Gaussian fluctuations, namely, the $\alpha$-stable L\'evy noise, can be characterized by the inspection of the switches from the superconducting to the resistive 
state of a JJ. In this case, the interesting information content can be effectively retrieved from the cumulative distribution functions of the switching currents~\cite{Gua19}.

In this work we shall deal with the characterization of the L\'evy noise sources, which correspond to stochastic processes that exhibit very long {\it distance} in a single displacement, namely, 
a {\it flight}. Results on the dynamics of systems driven by L\'evy flights have been reviewed in Refs.~\cite{Dub08, Zab15}. L\'evy noise, a generalization~\cite{Dub05b} of 
the Gaussian noise source~\cite{Val06,Val08,Val15,Fal13}, can be invoked to describe transport phenomena in different natural phenomena~\cite{Kad01,Sca03}, interdisciplinary applications~\cite{Dub08b,LaC10}, various condensed matter systems and practical applications. 
For instance, in graphene the presence of L\'evy distributed fluctuations has been recently discussed~\cite{Bri14,Gat16,Kis19}. In Ref.~\cite{Gua17}, it has been speculated that the anomalous premature switches affecting the switching currents in graphene-based JJs, that are likely to be unrelated to thermal fluctuations~\cite{Cos12}, could be ascribed to L\'evy distributed phenomena.
L\'evy processes emerge in the electron transport~\cite{Nov05} and optical properties~\cite{Shi01,Bro03} of semiconducting nanocrystals quantum dots, and also in photoluminescence experiments in moderately doped $n$-InP samples~\cite{Lur12,Sub14}. The L\'evy statistics was also invoked to model interstellar scintillations~\cite{Bol03,Bol05,Bol06,Gwi07} and the quasiballistic heat conduction in semiconductor alloys~\cite{Moh15,Upa16,Anu18}. 
On a more applicative and engineering side, L\'evy noise often appears in telecommunications and networks~\cite{Yan03,Bha06,Cor10} and has been also used to describe vibration data in industrial bearings~\cite{LiYu10,Cho14,Saa15} and in wind turbines rotation parts~\cite{Ely16}. In addition, L\'evy processes find applications in the mathematical modelling of random search processes as well as computer algorithms~\cite{Pal19}.
Therefore, a reliable device capable of detecting fluctuations distributed according to L\'evy statistics, of a signal that can be trasduced in a bias current feeded to the JJ, may be suitable in different frameworks.

Effects induced by non-Gaussian, \emph{i.e.}, L\'evy, distributed fluctuations 
have already been studied thoroughly in both short~\cite{Aug10,Spa12} and long~\cite{Gua13,Val14,Spa15,Gua15,Gua16} JJs. Notably, in the long junction case, the interplay between L\'evy 
and thermal noise and the generation of solitons~\cite{Par78,Ust98} was also investigated~\cite{Val14,Gua16,GuaGia16}. The aforementioned works calculate the mean first-passage time 
and the nonlinear relaxation time in short and long JJs, respectively, to deal with the ``premature'' switches, driven by L\'evy flights, from the superconducting metastable state. 
Also the escape of a particle from a trapping potential has been addressed~\cite{Che07}, as well as the average velocity of a particle in a washboard potential subject to noise has been 
discussed, for the diffusion problem~\cite{Sokolov09,Goychuk11} for {\it tempered}, \emph{i.e.}, truncated, L\'evy distributions~\cite{Gajda10}, in a tilted potential~\cite{Liu19}.

At variance with the analysis of the currents at which an underdamped junction switches to the finite voltage, the proposed noise detector is based on the measurement of the average 
voltage drop across an overdamped JJ biased by a constant electric current.
The rationale is that the voltage is thus proportional to the average speed, or mobility, of the biased JJ, that amounts to the speed of a particle in a tilted washboard potential under the effect 
of noise. The relation between the average velocity of a particle in these conditions and the features of the L\'evy noise is of the power-law type~\cite{Che07,Liu19}. It is therefore tempting 
to exploit the relation between the noise intensity and the voltage to infer the noise characteristics. We demonstrate that the proposed detection method for L\'evy-distributed fluctuations 
conveniently works at small bias currents and low temperatures, where switching processes due to thermal fluctuations as well as quantum tunneling \cite{Cas96,Longobardi11} can be neglected. Indeed, our proposal paves the way to the direct experimental investigation of an $\alpha$-stable L\'evy noise signal.

In this work, the characterization of the statistical fluctuations of the voltage in the JJs is proposed to discriminate the features of the noise affecting the device.
In the proposed set-up, it is assumed the presence of a L\'evy noise source, together with an intrinsic Gaussian thermal noise. 
Indeed, this detection (or discrimination) scheme is different from the classic acceptation in the statistical detection theory, where one usually supposes that the quantity to be revealed is inextricably mixed with noise.
The proposed device proves useful in the case of very weak signals at very high frequency, \emph{i.e.}, it is an alternative method to the standard electronics when the latter does not allow efficient and low noise sampling.

The paper is organized as follows. In Sec.~\ref{Sec01}, we discuss the operating principles of a Josephson-based noise detector: Sec.~\ref{Sec02a} lays the theoretical groundwork for 
the phase evolution of a short JJ and Sec.~\ref{Sec02b} describes the statistical properties of the L\'evy noise and the method employed for the stochastic simulations. In Sec.~\ref{Sec03}, the 
results are shown and analyzed. In Sec.~\ref{Sec03a} we compare the average voltage drop obtained numerically with the analytical estimate of the average velocity of the phase particle, in 
the case of an escape process driven by L\'evy flights from a washboard-potential well. We also investigate in Sec.~\ref{Sec04} the effects of the temperature at which the junction operates as 
a detector. In Sec.~\ref{Sec05}, conclusions are drawn.
\vspace{-0.5cm}
\section{Noise detector operating principles and Model}
\label{Sec01}\vskip-0.2cm

\begin{figure}[t!!]
\centering
\includegraphics[width=\columnwidth]{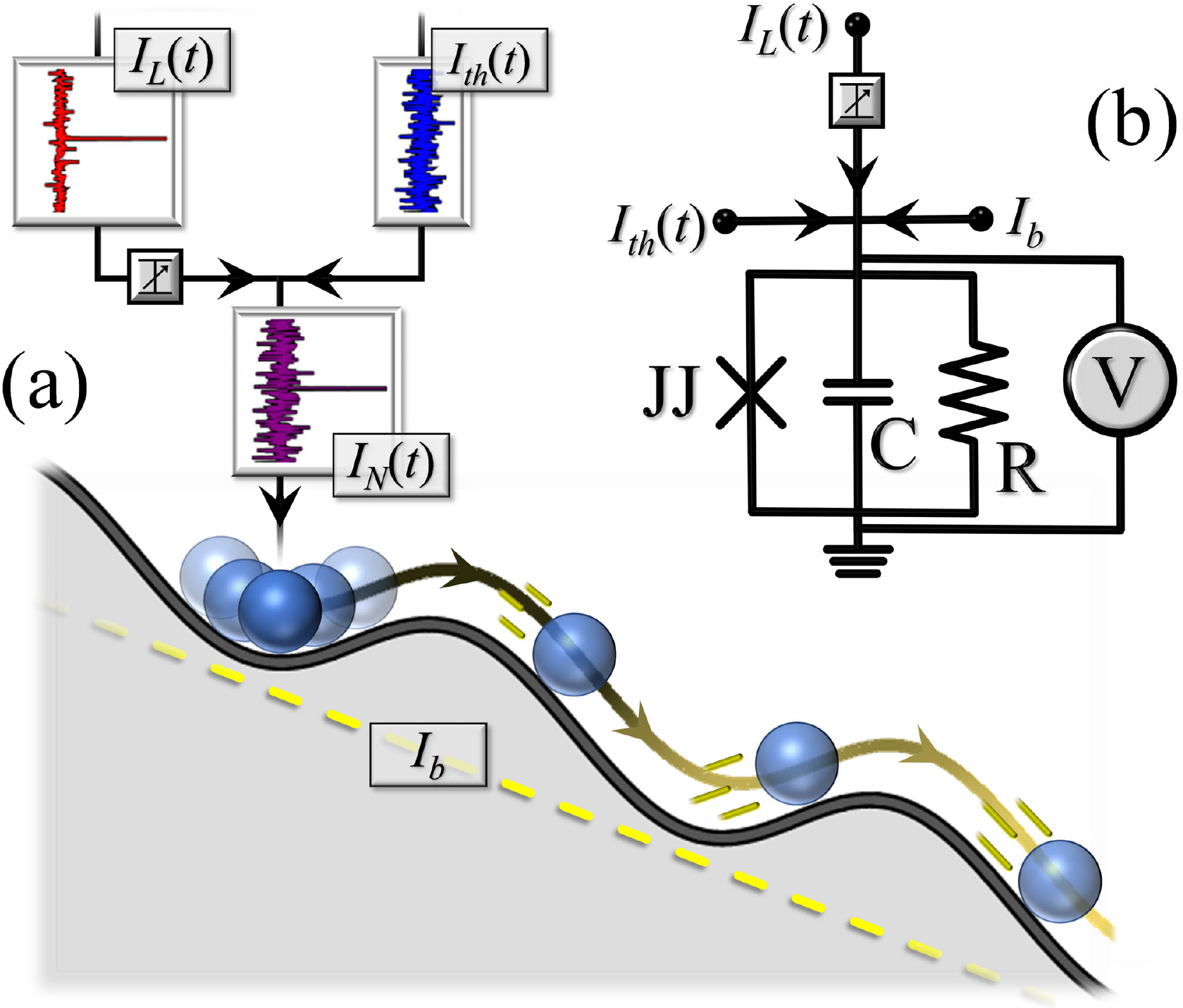}
\caption{(a) Phase particle in a potential minimum of the washboard potential $U$, in the case of a non-zero bias current that tilts the potential. The phase can overcome a potential barrier, 
rolling down along the potential, for the effect of the noise current, $I_N(t)$, which is the sum of thermal and L\'evy noisy contributions. (b) Simplified equivalent circuit diagram for the resistively 
and capacitively shunted junction model. The bias current, $I_b$, and the noise currents, $I_L(t)$ and $I_{th}(t)$, are included in the diagram.}
\label{Fig01}
\end{figure}

A setup for a Josephson based noise readout~\cite{Pek05,Suk07,Urb09,Gua19} consists of a JJ fed by two electric currents, $I_b$ and $I_N$. Specifically, $I_b$ is the bias current drawn from a parallel source and $I_N$ is the stochastic noise current, whose characteristics we wish to unveil. To this purpose we discuss a detection scheme based on the measurement of the average voltage drop across the junction. In our approach, the injected bias current is fixed at a value lower than the critical current, to steady keep the system in the superconducting metastable state until the noise eventually pushes out it, thus inducing a passage from the zero-voltage state to the finite voltage ``running'' state. 
In fact, the voltage in a JJ is proportional to the time derivative of the phase difference $\varphi$ between the wave functions describing the superconducting condensate in the two electrodes 
according to the a.c. Josephson relation $V =(\Phi _0 /2\pi) d\varphi /dt $~\cite{Jos62,Jos74}, where $\Phi _0=h/2e\simeq 2.067* 10^{-15}\;\text{V}\,\text{s}$ is the magnetic flux quantum. We seek for an analysis of the average voltage drop which allows to catch some features of the noise source affecting the phase dynamics.

In the experiments, we can reasonably expect that the amplitude of the current noise fluctuations is not precisely known. Since this noisy external signal is sent to the junction through 
an electric current, the amplitude of the fluctuations can be varied through an attenuator. This allows to experimentally measure the average voltage drop in correspondence of few different noise intensities. In this way, we demonstrate that it is possible to effectively evaluate the parameter $\alpha$, which characterizes the noise signal that perturbs the system, directly from the analysis of the average voltage across the junction. 

A natural issue in this procedure concerns how to change (attenuate) the amplitude of L\'evy current contribution. 
To reduce the current in a controlled way it might be convenient to use a cryogenic delay line (transmission line) in the small loss regime, under the Heaviside condition. 
With such a delay line, the signal output of the attenuator can be suitably weakened by varying the transmission line length. 
The Heaviside condition should be valid in all the frequency-band of L\'evy noise, to ensure that the signal is less distorted while propagating in the transmission line.
In practice, some cut-offs due to the physics of the problem reduce the band of interest. 
In the present set-up, the voltage device is measured within some time interval, \emph{i.e.}, $\tau_{\text{max}}$, thus one can assume that the spectrum is negligible below $f = 1/\tau_{\text{max}}$.
Moreover, JJs do not respond to frequencies that are much larger than the resonant frequency $\omega_J$, and surely below the frequency at which the Cooper pairs are broken, that is $h f \le \Delta$, where $\Delta$ is the superconductor gap.
So, it suffices that the transmission line does not distort the input noise in the bandwidth $1/\tau_{\text{max}} \le f \le \Delta/h$.


\subsection{The model}
\label{Sec02a}\vskip-0.2cm

A tunnel Josephson junction is a quantum device formed by sandwiching a thin insulating layer between two superconducting electrodes. In the following, we 
consider a short JJ, in which the physical length of the junction is lower than the characteristic length scale of the system, that is the Josephson penetration length, 
$\lambda_{_{J}} = \sqrt{\frac{\Phi_0}{2\pi \mu_0}\frac{1}{t_d J_c}}$~\cite{Bar82}. Here, $t_d=\lambda_{L,1}+\lambda_{L,2}+d$ is the effective magnetic thickness, with $\lambda_{L,i}$ and $d$ being the London penetration depth of the $i$-th electrodes and the insulating layer thickness, respectively, $\mu_0$ 
is the vacuum permeability, and $J_c$ is the critical current area density. To give a realistic estimate of this length scale, let us consider, for instance, a Nb/AlO/Nb 
junction with a normal-state resistance per area $R_a \approx 50\;\Omega \left( \mu\text{m} \right)^2$, a low-temperatures critical current area density equal to 
$J_c = \frac{\pi}{2} \frac{1.764k_BT_c}{eR_a} \approx 40 \;\mu\text{A} / \left( \mu\text{m}\right)^2$~\cite{Bar82}, and the effective magnetic thickness 
$t_d \approx 160 \;\text{nm}$, assuming $T_c = 9.2\;\text{K}$ and $\lambda_{L}^0 \approx 80\;\text{nm}$ for Nb. With these parameter values, the Josephson penetration depth reads $\lJ \simeq 6 \;\mu\text{m}$. A {\it short} Josephson tunnel junction is a junction in which both lateral dimensions $\nL$ and $\nW$ 
are lower than the Josephson penetration depth, $\lambda_{_{J}}$. The dynamics of the Josephson phase $\varphi$ for a dissipative, current-biased short JJ can be 
studied within the resistively and capacitively shunted junction (RCSJ) model~\cite{Bar82,GuaVal15,Spa17} according to the following equation
\begin{equation}
\left ( \frac{\Phi_0}{2\pi} \right )^{\!\!2} C \frac{d^2 \varphi}{d \tau^2}+\left ( \frac{\Phi_0}{2\pi} \right )^{\!\!2}\frac{1}{R} \frac{d \varphi}{d \tau}+\frac{d }{d \varphi}U 
= \left ( \frac{\Phi_0}{2\pi} \right ) I_N.
\label{RCSJ}
\end{equation}
The coefficients $R$ and $C$ are the normal-state resistance and the capacitance of the JJ, respectively, and $U$ is the washboard potential along which the phase evolves,
\begin{equation}
U(\varphi,i_b)=E_{J_0}\left [1- \cos(\varphi) -i_b\varphi\right ],
\label{Washboard}
\end{equation}
where $E_{J_0}=\left ( \Phi_0/2\pi \right )I_c$ and $i_b$ is the bias current normalized to the critical current $I_c$. The resulting activation energy barrier, $\Delta U(i_b)$, 
confines the phase $\varphi$ in a metastable potential minimum and can be calculated as the difference between the maximum and minimum value of $U(\varphi,i_b)$. 
In units of $E_{J_0}$, it can be expressed as
\begin{equation}
\Delta \mathcal{U}(i_b)=\frac{\Delta U(i_b)}{E_{J_0}}=2\left [ \sqrt{1-i_b^2} -i_b\arcsin(i_b)\right ].
\label{activationenergybarrier}
\end{equation}
In the phase particle picture, the term $i_b$ represents the tilting of the potential profile, see Fig.\ref{Fig01}; with increasing $i_b$ the slope of the washboard increases and the height $\Delta \mathcal{U}(i_b)$ of the right potential barrier reduces, until it vanishes when $i_b=1$, that is when the bias current coincides with the critical value. 

If one normalizes the time to the inverse of the characteristic frequency, that is $t = \tau \omega_c$ with $\omega_c=\left ( 2e/\hbar \right )I_cR$, Eq.~\eqref{RCSJ} 
can be cast in the dimensionless form
\begin{equation}
\bC\frac{d^2 \varphi (t)}{dt^2}+ \frac{d \varphi (t)}{dt} + \sin \left [ \varphi\left ( t \right ) \right ] = i_{N}(t) + i_b,
\label{RCSJnorm}
\end{equation}
where $\bC=\omega_c RC$ is the Stewart-McCumber parameter. A highly damped (or overdamped) junction has $\bC\ll 1$, that is, in other words, a small capacitance and/or 
a small resistance. In contrast, a junction with $\bC\gg 1$ has large capacitance and/or large resistance, and is weakly damped (or underdamped).

\subsection{The statistical model}
\label{Sec02b}\vskip-0.2cm

Equation~\eqref{RCSJnorm} balances the three contributions the Josephson elements on the left side, \emph{i.e.}, the capacitive term, the dissipative contribution, and the Josephson supercurrent, with the two terms on the right side, \emph{i.e.}, the external bias current $i_b= I_b / I_c$ and the current noise $i_N (t) = I_N (t ) / I_c$. In this work, the random current is modeled as a mixture of a standard Gaussian white noise, associated to the JJ resistance, and a stochastic L\'evy 
process. This current is modeled with the approximated finite independent increments~\cite{Chechkin00}.
If we consider both Gaussian and L\'evy-distributed fluctuations, with amplitudes $\gG$ and $\gL$, respectively, the stochastic independent increment reads
\begin{equation}
\Delta i_N \simeq 
 \sqrt{ 2\gG \Delta t\; }\; N\left(0,1 \right) + 
 \left( \gL \Delta t \right) ^{1/\alpha}S_{\alpha}\left(1,0,0\right).
\label{GLFincr}
\end{equation}
Here, the symbol $N\left(0, 1 \right)$ indicates a random function Gaussianly distributed with zero mean and unit standard deviation, while $S_{\alpha}(1, 0, 0)$ denotes a \emph{standard} 
$\alpha$-stable random L\'evy variable. 
For the sake of clarity, we briefly review the concept of $\alpha$-stable L\'evy distributions~\cite{Ber96,Sat99,Gne54,Khi36,Khi38,Fel71}. A random non-degenerate variable $X$ is stable if
\begin{equation}
\forall n\in\mathbb{N}, \exists (a_n,b_n) \in \mathbb{R}^+\times\mathbb{R}: \quad X+b_n=a_n\sum_{j=1}^{n} X_j, \label{AlfaStable}
\end{equation}
where the $X_j$ terms are independent copies of $X$. Besides, $X$ is strictly stable if, and only if, $b_n=0 \,\,\, \forall n$. The Gaussian distribution belongs to this class. The definition of characteristic function for a random variable $X$ with an associated distribution function $F(x)$ is
\begin{equation}\label{GenerealCharFunc}
\phi (u) = \left < e^{iuX} \right > = \int_{-\infty}^{+\infty}e^{iuX}dF(x).
\end{equation}
Accordingly, a random variable $X$ is said to be stable if, and only if,
\begin{equation}
\exists (\alpha ,\sigma, \beta, \mu )\in(0, 2]\times \mathbb{R}^{+}\times [-1, 1]\times \mathbb{R}: \;\; X\overset{d}{=}\sigma Z+\mu,
\label{XStableFunc}
\end{equation}
with $Z$ being a random variable with characteristic function
\begin{eqnarray}\label{XCharFunc}
\phi(u)=\left\{\begin{matrix}
\exp \left \{-\left | u \right |^\alpha\left [ 1-i\beta \tan\frac{\pi\alpha}{2}(\textup{sign}u) \right ]\right \} \; \alpha\neq 1\\
\exp \left \{-\left | u \right |\left [ 1+i\beta\frac{2}{\pi}(\textup{sign}u)\log \left | u \right |\right ]\right \} \; \alpha=1
\end{matrix}\right. \qquad
\end{eqnarray}
in which $\textup{sign}u=0$ for $u=0$ and $\textup{sign}u\pm1$ for $u\gtrless0$.
These distributions are symmetric around zero when $\beta=0$ and $\mu=0$. In Eq.~(\ref{XCharFunc}) for the $\alpha=1$ case, $0\cdot \log0$ is always interpreted as $\lim_{x\to0} x\log x=0$, giving $\phi(0)=1$.

In general, the notation $S_{\alpha}(\sigma, \beta, \lambda)$ is used for indicating L\'evy distributions~\cite{Gua13,Val14,Spa15,Gua15,Gua16}, 
where $\alpha\in(0,2]$ is the {\it stability index}, $\beta\in[-1,1]$ is the {\it asymmetry parameter}, and $\sigma>0$ and $\lambda$ are the scale and location parameters,
respectively. The stability index characterizes the asymptotic long-tail power law for the distribution, which for $\alpha<2$ is of the $\left | x \right |^{-\left ( 1+\alpha \right )}$ type. 
The case $\alpha=2$ is the Gaussian distribution. In fact, the probability density function of a normal distribution $N\left(\lambda, \sigma \right)$ is that of the stable distribution $S_{2}(\sigma/\sqrt{2}, \beta, \lambda)$. In this work, we consider symmetric (\emph{i.e.}, $\beta=0$), bell-shaped, standard (\emph{i.e.}, with $\sigma=1$ and $\lambda=0$), 
stable distributions $S_{\alpha}(1, 0, 0)$, with $\alpha\in[0.1,2]$. 
A physical interpretation of L\'evy fluctuations can be inferred from the 
understanding of the structure of the paths of L\'evy processes. Indeed, a linear combination of a finite number of independent L\'evy processes is again a L\'evy process. It turns out that one may consider any L\'evy process as an independent sum of a Brownian motion with drift and a countable number of independent Poisson processes with different jump rates, jump distributions, and drifts. 
This is the L\'evy-It\^o decomposition theorem, see Ref.~\cite{Applebaum09} and references therein. 
To simulate the L\'evy noise sources it has been used the algorithm proposed by Weron~\cite{Wer96} to implement the Chambers method~\cite{Cha76}. The stochastic integration of Eq.~\eqref{RCSJnorm} is performed 
with a finite-difference explicit method, using a time integration step $\Delta t=10^{-2}$. 

It might be useful to give some physical considerations on the parameter $\gG$ in Eq.~\eqref{GLFincr}. In the pure Gaussian noise case, \emph{i.e.}, $\gL=0$ so that 
$I_N\equiv I_{th}$, the statistical properties of the current fluctuations, in physical units, are given by
\begin{eqnarray}
\textup{E}\left [ I_{th}\left ( \tau\right ) \right ] &=& 0 \nonumber \\
\textup{E}\left [ I_{th}\left (\tau\right)I_{th}\left (\tau+\tilde{\tau}\right) \right ] &=& 2\frac{k_BT}{R}\delta \left (\tilde{\tau} \right ), 
\label{WNProperties}
\end{eqnarray}
where $\textup{E}[\cdot ]$ is the expectation operator. In our normalized units, the same equations become 
\begin{eqnarray}
\textup{E}\left [ i_{th}(t) \right ] &=& 0, \nonumber \\
\textup{E}\left [ i_{th}(t)i_{th}\left (t+\tilde{t}\,\right) \right ] &=& 4\gamma_{G} (T)\delta \left (\tilde{t} \,\right),
\label{WNCorrelation}
\end{eqnarray}
where the amplitude of the normalized correlator is connected to the physical temperature through the relation
\begin{eqnarray}
\label{WNAmp}
\gG (T)= \frac{k_BT}{2R}\frac{\omega_c}{I^2_c}=\frac{k_BT}{2E_{J_0}}.
\end{eqnarray}
It is worth stressing that, with the time normalization used in this work, the noise intensity $\gG$ can be expressed as the ratio between the thermal energy and the Josephson 
coupling energy $E_{J_0}$. As usual for numerical simulations in normalized units, the reported quantities, as the Gaussian noise amplitude, should be related to physical 
quantities through the system physical parameters, \emph{e.g.}, the critical current, the normal resistance, the capacitance, and the temperature of the device. For instance, 
for a junction with a critical current $I_c = 1\;\mu \textup{A}$ at a temperature $T= 0.5\;\textup{K}$ the dimensionless noise amplitude is $\gG\sim10^{-2}$.

The detection method proposed in this work is based on the measurement of the average voltage drop across the junction. Here the average is intended as a double 
averaging, that is ensemble and time averages. In the \textit{i}-th numerical realization, the time average of the voltage difference across the JJ 
can be obtained as follows
\begin{eqnarray}\label{MeanV_i}\nonumber
\left \langle V_i \right \rangle&=&\frac{1}{\tau_{\text{max}}}\int_{0}^{\tau_{\text{max}}}\frac{\Phi_0}{2\pi}\frac{\mathrm{d} \varphi_i \left ( \tau \right )}{\mathrm{d} \tau}d\tau\\
&=&\frac{\Phi_0\omega_c}{2\pi}\frac{\varphi_i(t_{\text{max}})-\arcsin (i_b)}{t_{\text{max}}},
\end{eqnarray}
$\varphi(0)=\arcsin (i_b)$ being the initial phase and $t_{\text{max}}=\omega_c\tau_{\text{max}}$ the normalized measurement time. The average voltage drop across the 
junction is finally obtained by averaging over the total number of independent numerical repetitions $N_{\text{exp}}$. In units of $\Phi_0\omega_c$, it reads
\begin{equation}\label{MeanV}
\V=\frac{\left \langle V \right \rangle}{\Phi_0\omega_c}=\frac{1}{N_{\text{exp}}}\sum_{i=1}^{N_{\text{exp}}}\frac{\left \langle V_i \right \rangle}{\Phi_0\omega_c}.
\end{equation}
In the following, the value of $\V$ is estimated averaging over a normalized time $t_{\text{max}}=10^4$ and a number of independent numerical repetitions $N_{\text{exp}}=10^4$. 
\begin{figure}[t!!]
\centering
\includegraphics[width=\columnwidth]{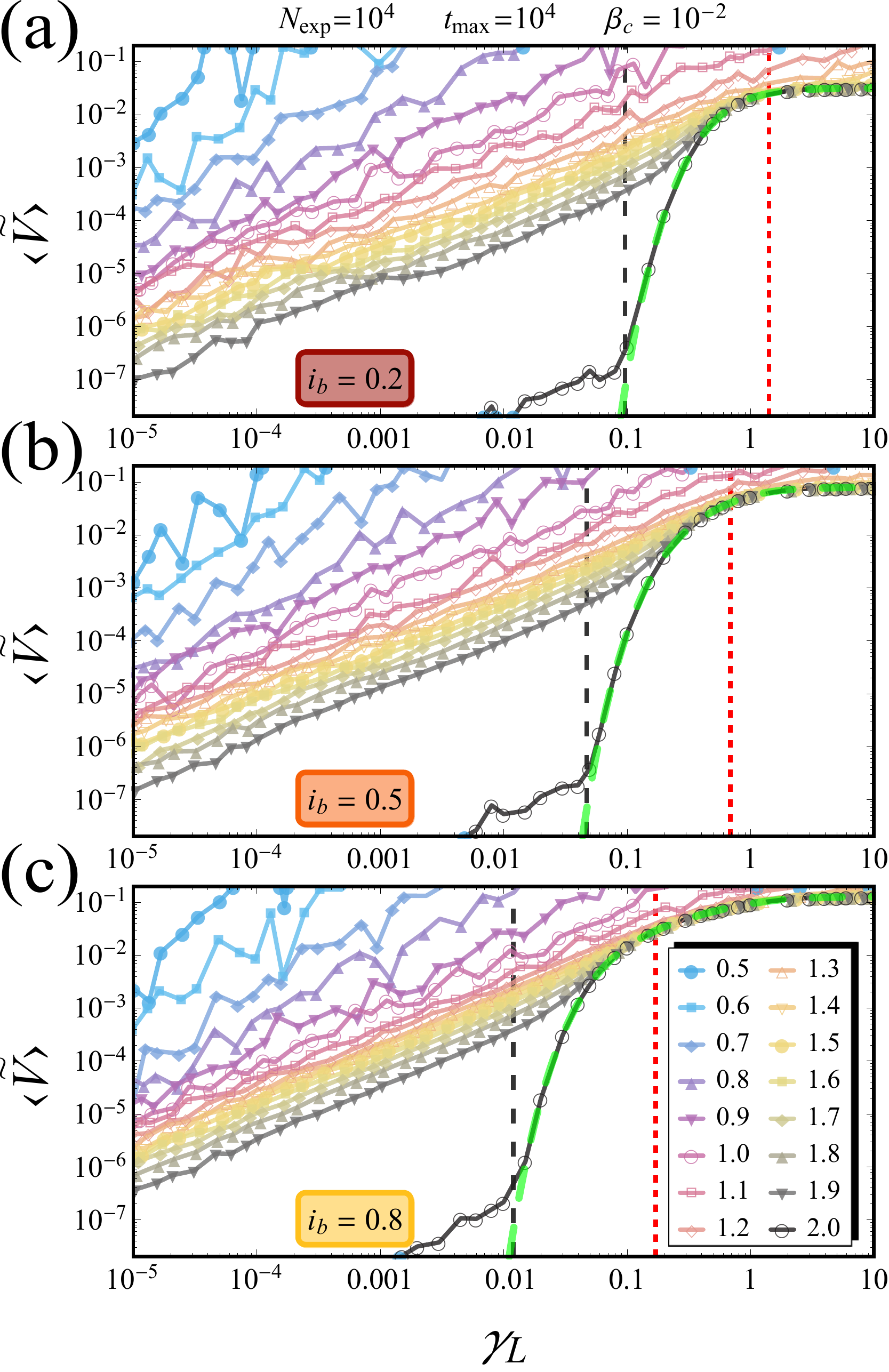}
\caption{Normalized average voltage drop $\V$ as a function of the L\'evy noise intensity $\gL$, for different values of the parameter $\alpha$ and at different bias currents 
$i_b=\{0.2, 0.5,\text{ and } 0.8\}$, see panels (a), (b), and (c), respectively. The red short-dashed line indicates the noise amplitude that equals the activation energy barrier 
$\gL = \Delta \mathcal{U}(i_b)$, see Eq.~\eqref{activationenergybarrier}, while the black long-dashed line indicates the noise amplitude $\gamma_{_L}^{th} $ at which the 
inverse Kramers rate, see Eq.~\eqref{Kramersescape}, equals the measurement time $t_{\text{max}}=10^4$. The green dashed curve in all panels indicates the average voltage drop versus the noise intensity, for thermal fluctuations, analytically calculated in Refs.~\cite{Lee71,Bar82}. Lines in the figure are guides for the eye. Legend in panel (c) 
refers to all panels.}
\label{Fig02}
\end{figure}

\section{Results and discussions}
\label{Sec03}\vskip-0.2cm

In the overdamped ($\beta_C=0.01$) junction case here considered, we initially neglect Gaussian thermal fluctuations ($\gG=0$) to emphasize the influence of L\'evy flights. 
The Gaussian noise source will be taken into account at a later stage, to explore the robustness of the detection through I-V analysis in the presence of thermal noise.

In Fig.~\ref{Fig02} we illustrate the behavior of the normalized average voltage drop $\V$ as a function of the L\'evy noise intensity $\gL$, for several values of the stability 
index $\alpha$, and three different bias current points, $i_b=\{0.2,0.5, \text{ and }0.8\}$, see panels (a), (b), and (c), respectively. In the top panel of Fig.~\ref{Fig02} obtained 
for $i_b=0.2$, interestingly, for $\gL$ values below a threshold marked with a red short-dashed vertical line, the $\V$ vs $\gL$ curves look quite similar: in fact, changing the 
index $\alpha$, the average voltage data are arranged in well-distinct parallel lines (in the log-log scale) with a positive slope. The aforementioned 
threshold is given by the activation energy barrier, $\Delta \mathcal{U}(i_b)$. This means that, for noise amplitudes lower than the activation energy barrier, 
$\gL<\Delta \mathcal{U}(i_b)$, the $\V$ curves follow a power law behavior \cite{Che07} of the $\tV \times \gamma_{_L}^{\mu_{\alpha}}$ type with an exponent 
$\mu_{\alpha} \simeq 1$. 

The curve for $\alpha = 2$ is an exception, since in this case the L\'evy distribution amounts to the Gaussian case; L\'evy flights are indeed 
missing and the $\V$ curve is several orders of magnitude lower than the $\alpha=1.9$. Anyway, also the curve for $\alpha = 2$ shows two distinct behaviors, respectively 
above and below a certain threshold that is highlighted in Fig.~\ref{Fig02} with a black vertical long-dashed line.
This threshold can be estimated as the noise intensity at which the inverse Kramers rate~\cite{Kra40} matches the measurement time, that is $\gamma_{_L}^{th}\equiv\sqrt{2}\gG$ 
so that $r(i_b,\gG)^{-1}=\tau_{\text{max}}$, where the coefficient $\sqrt{2}$ stems from the different normalization of the noise amplitudes of a Gaussian and a L\'evy 
distribution with $\alpha=2$. According to the Kramers theory, the escape rate from a confining barrier, see Eq.~\eqref{activationenergybarrier}, reads
\begin {equation}
\label{Kramersescape}
r(i_b,\gG)=\frac{\omega_A}{2\pi}e^{-\frac{\Delta U(i_b)}{ k_BT}}=\frac{\omega_c}{2\pi}\left ( 1-i_{b}^{2} \right )^{\frac{1}{4}}e^{-\frac{\Delta \mathcal{U}(i_b)}{2 \gG}},
\end {equation}
which is obtained assuming the strong damping limit for the attempt jump frequency, $\omega_A=\omega_c\left ( 1-i_{b}^{2} \right )^{\frac{1}{4}}$, and a noise amplitude given 
by Eq.~\eqref{WNAmp}. Thus, Fig.~\ref{Fig02} demonstrates that at low noise amplitudes the Gaussian distributed fluctuations are not intense enough to induce escapes in the 
measurement time $t_{\text{max}}$. To put it in another way, for $\alpha=2$ at noise intensities $\gL<\gamma_{_L}^{th}$ the phase particle remains confined within the initial state, 
and therefore the values of $\V$ are vanishingly small. Conversely, for higher intensities, $\gL>\gamma_{_L}^{th}$, noise-induced switches can be triggered. In this case, the phase 
particle can leave the initial metastable state rolling down along the washboard potential. The speed of the phase particle therefore increases and a non-negligible average voltage 
drop appears. In this case, the curve obtained numerically, for $\gL>\gamma_{_L}^{th}$, perfectly matches the average voltage drop analytically calculated for the 
case of a finite junction capacitance and in the presence of thermal fluctuation, see the analytical expression derived in Refs.~\cite{Lee71,Bar82} and reported in~\footnote{Following the same mathematical procedure discussed in Ref.~\cite{Bar82}, we can evaluate the current-voltage characteristic, amended to include the first order effects of a finite capacitance:
\begin{eqnarray}\label{a1}
<V>&=&\frac{2}{\gamma}R_{N}I_{c}\frac{\exp(\pi \gamma \alpha)-1}{\exp(\pi \gamma \alpha)} T^{-1}_{1} \left ( 1+\Omega^2 \frac{T_{2}}{T_{1}} \right )\\\label{a2}
T_{1}&=&\int_{0}^{2\pi}\!\!d\varphi \,\text{I}_{0}\bigg(\gamma \sin\frac{\varphi}{2}\bigg) \exp\left [ -\frac{\gamma }{2}\alpha\varphi \right ]\\\label{a3}
T_{2}&=&\int_{0}^{2\pi}\!\!d\varphi \sin\left ( \frac{\varphi}{2} \right ) \text{I}_{1}\left ( \gamma \sin\frac{\varphi}{2} \right )\exp\left [ -\frac{\gamma }{2}\alpha\varphi \right ],
\end{eqnarray}
where $\text{I}_0(x)$ and $\text{I}_1(x)$ are modified Bessel functions. 
Since we prefer to write these equations in the same notation as Ref.~\cite{Bar82}, to help the reader we show a comparison between the notation used in this work and that of Ref.~\cite{Bar82}: $\Omega\equiv\sqrt{\beta_C}$, $\alpha\equiv i_b$, and $\gamma\equiv1/\gamma_G$.}, which is indicated by the green-dashed curve in Fig.~\ref{Fig02}. The discrepancies shown in Fig.~\ref{Fig02} for 
$\gL < \gamma_{_L}^{th}$ are ascribable to the finite measurement time. For longer computational, \emph{i.e.}, measurement, time these discrepancies tend to disappear and the matching with the analytical expression improves considerably. 

A further increase of the noise intensity bears little consequences, once the fluctuations are intense enough to overcome the potential barrier. This is why, for $\gL>\Delta U(i_b)$, 
the $\V$ curves for different $\alpha$ values tend to a common plateau.

The overall scenario described so far essentially persists with increasing $i_b$, see panels (b) and (c) of Fig.~\ref{Fig02} for $i_b=0.5$ and $i_b=0.8$, respectively. 
However, some differences come to light in agreement with Eq.~(\ref{Kramersescape}). 
In fact, at large bias current the potential is increasingly tilted and the activation energy barrier decreases; this is the reason why the thresholds marked by vertical dashed lines move 
leftwards, the $\V$ curves shift towards lower $\gL$ values, and the linear trend (in the log-log scale) appears at lower $\gL$ values. Moreover, the spacing between these curves 
reduces while increasing $i_b$. Finally, the value approached by $\V$ at noise amplitudes beyond the barrier energy threshold, \emph{i.e.} for $\gL>\Delta U(i_b)$, increases with $i_b$.

The linear portion of the $\V$ vs $\gL$ curves essentially embodies the detection features we are interested in. In this region, all curves of Fig.~\ref{Fig02} can be fitted with 
the function $\tV\times \gamma_{_L}^{\mu_{\alpha}}$, $\tV$ being the fitting parameter and $\mu_{\alpha} \simeq 1$ for the L\'evy noise escapes \cite{Che07}. Thus, by ranging the 
noise amplitude in a suitable interval, from an estimate of the fitting parameter $\tV$ we can infer the value of the stability index $\alpha$. We note that the increasing 
fluctuations shown by the curves of $\V$ vs $\gL$ for $\alpha \lesssim 1$ are due to the finite value chosen for the measurement time. Indeed, the average behavior of all these curves shows a power-law trend, with well-distinct parallel lines in the log-log scale, and these fluctuations tend to be smoothed out by increasing the measurement time. 

\begin{figure}[t!!]
\centering
\includegraphics[width=\columnwidth]{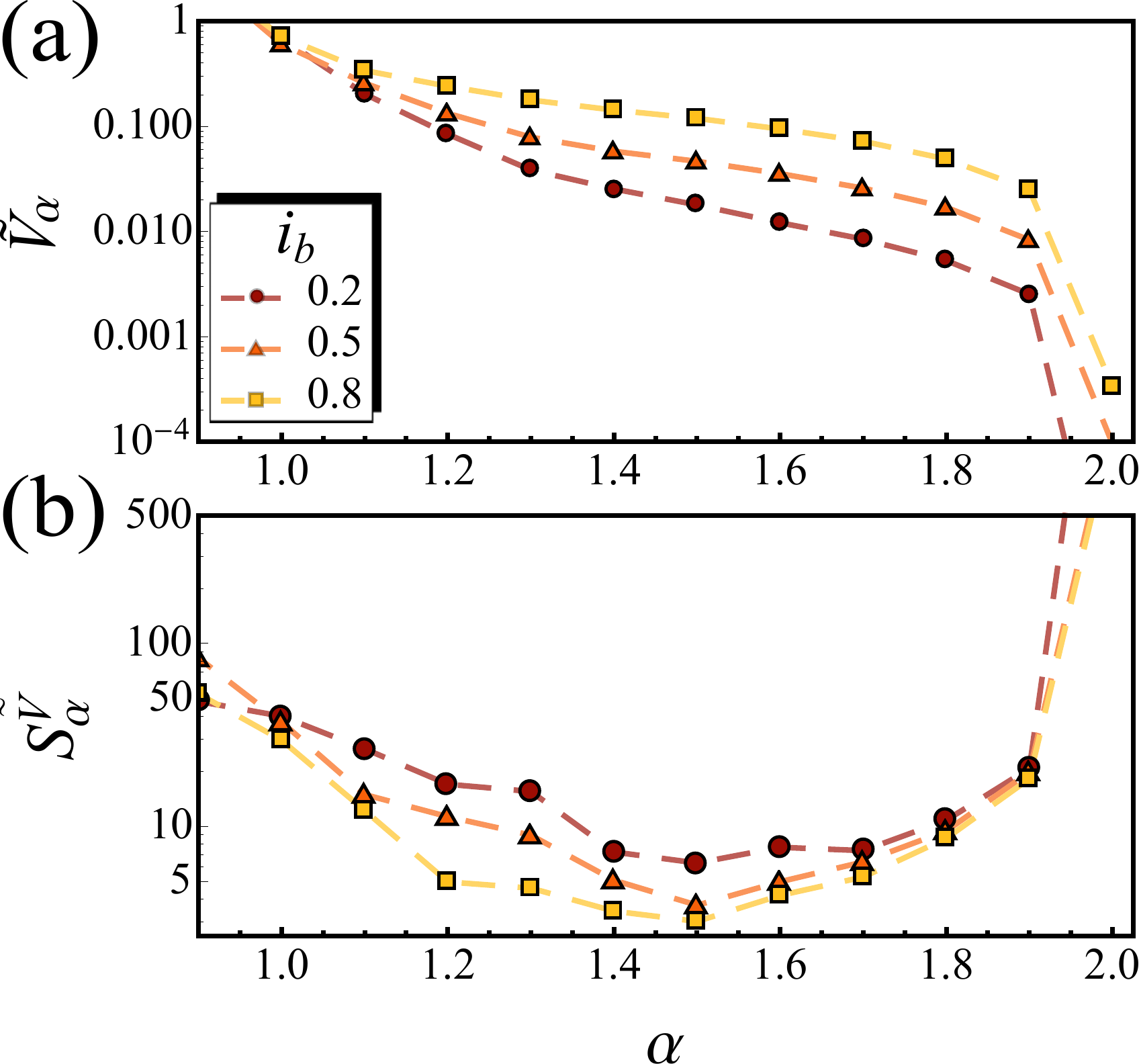}
\caption{(a) Fitting parameter $\tV$ versus $\alpha$ at different $i_b$ values. 
The parameter $\tV$ is obtained fitting the $\V$ vs $\gL$ curves shown in Fig.~\ref{Fig02} in the range $\gL\sim[10^{-4},10^{-2}]$ with the function $\V=\tV\times\gamma_{_L}$. 
(b) Sensitivity $\S$ as a function of $\alpha$ at different $i_b$ values. The dashed lines are guides for the eye. Legend in panel (a) refers to both panels.}\label{Fig03}
\end{figure}

Figure~\ref{Fig03}(a) shows the behavior of the fitting parameter $\tV$ as a function of the parameter $\alpha$, at different bias currents $i_b=\{0.2, 0.5,\text{ and } 0.8\}$, extracted 
from the Fig.~\ref{Fig02} in the range of noise amplitude $\gL\sim[10^{-4},10^{-2}]$. 

First, we observe that the fitting parameter $\tV$ monotonically reduces by increasing $\alpha$. This behavior confirms that, at a given bias current, an experimental measurement of 
$\tV$ returns the stability index $\alpha$. However, from Fig.~\ref{Fig03}(a) it is also clear that, at a given variation of $\alpha$, the fitting parameter $\tV$ changes more at a lower 
bias $i_b$. This means that a small bias current is favorable for the detection strategy. This feature is quantified by the relevant figure of merit of the detector, that is the voltage sensitivity, 
$\S$. This is defined as the ratio between the percentage variation of the voltage fitting parameter, $\tV$, and the percentage variation of the system parameter $\alpha$. Since we are 
considering $\alpha$ variations equal to $\Delta\alpha=0.1$, we can calculate the sensitivity as
\begin{equation}\label{MeanV}
\S=\frac{\alpha}{\tV}\left | \frac{\Delta\tV}{\Delta\alpha} \right |=10\,\alpha\left ( \frac{\widetilde{V}_{\alpha-1}}{\tV} -1\right ).
\end{equation}
The capability of the device to discern the presence of a L\'evy component by measuring the average voltage drop is higher when the sensitivity increases. In Fig.~\ref{Fig03}(b) 
we show the behavior of $\S$ as a function of the parameter $\alpha$, at different bias currents. The sensitivity behaves non-monotonically, showing a minimum at $\alpha=1.5$. 
Markedly, $\S$ is larger at a lower $i_b$, as expected. Interestingly, the fact that for $\alpha=2$ the sensitivity is orders of magnitude larger than that for $\alpha=1.9$ suggests 
that the detection method is quite effective to recognize the presence of any L\'evy noise component with respect to the pure Gaussian noise case. This can be qualitatively 
understood, because the effects of Gaussian noise become exponentially small when the noise intensity is below the energy barrier.

\subsection{Average speed in the presence of L\'evy flights}
\label{Sec03a}\vskip-0.2cm

In this section we demonstrate the connection between the linear behavior of the average voltage drop that emerges at intensities $\gL < \Delta \mathcal{U}(i_b)$, 
that is where the relation $\V=\tV\times \gamma_{_L}^{\mu_{\alpha}}$ holds, and the features of L\'evy driven escape processes from a metastable state of the washboard potential. 
In particular, we observe that the fitting parameter $\tV$ can be estimated recalling that the phase particle can undergo $2\pi$ jumps along the washboard potential and that the 
mean escape time for the L\'evy statistics follows a power-law asymptotic behaviour~\cite{Che05,Che07,Dub08,Gua19} 
\begin{equation}
\tau_{_L}\left ( \alpha,\gL \right ) = \left ( \frac{ \Delta x } {2} \right )^{\alpha} \frac{{\cal C}_{ \alpha}} {\gamma_{_L}^{\mu_{\alpha} }}.
\label{tau_Levy}
\end{equation}
The scaling exponent $\mu_{\alpha}\simeq1$ and the coefficient ${\cal C}_\alpha$ are supposed to have a universal behaviour for overdamped systems. The previous equation 
shows that, unlike the Kramers rate, in the case of L\'evy flights the mean escape time is independent on the barrier height $\Delta U$, but only depends upon the distance 
$\Delta x$ between a minimum and a maximum of the washboard potential.

We observe that the normalized average voltage drop in Eq.~\eqref{MeanV} represents essentially the average speed in the case of escape processes from a metastable state
\begin{equation}\label{averagespeed}
\left \langle v_i \right \rangle=\frac{1}{t_{\text{max}}}\frac{\varphi_i(t_{\text{max}})-\arcsin (i_b)}{2\pi}=\frac{{\cal N}_{\text{jump}}}{t_{\text{max}}},
\end{equation}
where ${\cal N}_{\text{jump}}$ indicates the number of $2\pi$-slips that the phase particle makes to reach, in the time $t_{\text{max}}$, the position $\varphi_i(t_{\text{max}})$, starting from the initial state $\varphi(0)=\arcsin (i_b)$. 

\begin{figure}[t!!]
\centering
\includegraphics[width=\columnwidth]{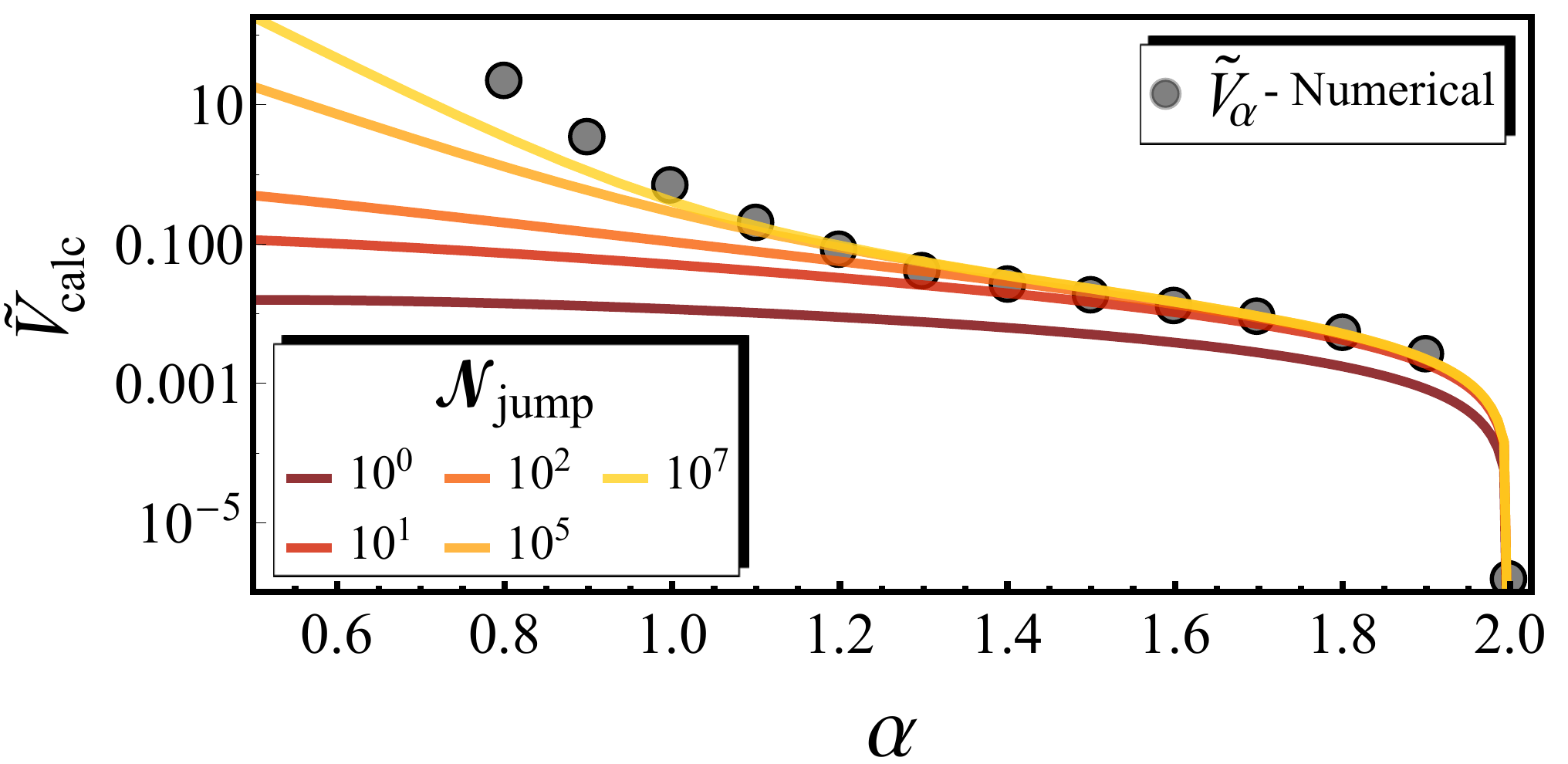}
\caption{Behavior of $\cV(\alpha,i_b,{\cal N}_{\text{jump}})$, see Eq.~\eqref{Vcalc}, as a function of $\alpha$ at different values of ${\cal N}_{\text{jump}}$ and $i_b=0.2$ (solid lines). 
For comparison, the numerical estimations of $\tV$ as a function of $\alpha$ at $i_b=0.2$ is also shown with gray circles.}\label{Fig04}
\end{figure}

Let us assume that the phase particle takes a time $\tau_{_L}$ to sweep ${\cal N}$ potential minima with a single jump; in this case, the average velocity can be estimated according 
to $\left \langle v_{_{{\cal N}}}\right \rangle=\frac{{\cal N}}{\tau_{_L}}=\left [\frac{{\cal N}}{{\cal C}_\alpha}\left ( \frac{2}{\Delta x} \right )^{\alpha} \right ]\times\gL$. Furthermore, the particle 
can leave the metastable well in which it resides moving to the left or to the right. Thus, the distances covered by a rightward ($\Delta x_l$) or a leftward ($\Delta x_r$) jump across 
${\cal N}$ minima can be calculated as $\Delta x_{r/l}(i_b,{\cal N})=(2{\cal N}+1)\pi\mp2\text{arcsin}(i_b)$. Finally, considering all possible jumps up to ${\cal N}_{\text{jump}}$, the 
average velocity can be estimated as $\left \langle v \right \rangle=\cV\times\gL$, where
\begin{eqnarray}\label{Vcalc}
&&\cV(\alpha,i_b,{\cal N}_{\text{jump}})=\\\nonumber
&&\sum_{{\cal N}=1}^{{\cal N}_{\text{jump}}}
\Bigg \{ \frac{{\cal N}}{{\cal C}_\alpha} \left \{ \left [ \frac{2}{\Delta x_{r}(i_b,{\cal N})} \right ]^{\alpha}- \left [ \frac{2}{\Delta x_{l}(i_b,{\cal N})} \right ]^{\alpha} \right \}\Bigg \}.
\end{eqnarray}
We note that the series in the previous equation, for ${\cal N}_{\text{jump}}\to+\infty$, converges only for $\alpha>1$, even if from a physical point of view the number of 
jumps can be quite large, but always finite, within a fixed observation time. In the following, for simplicity we assume for the coefficients ${\cal C}_\alpha$ the behavior given in Ref.~\cite{Che05}, namely, 
${\cal C}_\alpha=\Gamma (1-\alpha) \cos(\pi \alpha/2 )$, for an overdamped escape dynamics across a fixed height barrier of a cubic potential.

The behavior of $\cV(\alpha,i_b,{\cal N}_{\text{jump}})$ as a function of $\alpha$ at different values of ${\cal N}_{\text{jump}}$ and $i_b=0.2$ is shown in Fig.~\ref{Fig04}. 
For a prompt comparison with the numerical results shown in Fig.~\ref{Fig03}(a), we include also the behavior of the fitting parameter $\tV$ as a function of $\alpha$ at $i_b=0.2$. 
It is evident that the simple analytical estimate given in Eq.~\eqref{Vcalc} closely agrees with the numerical results for $\alpha\gtrsim1$, especially at a low $i_b$ value. However, we note that for $\alpha \lesssim 1$ we get a qualitative agreement between analytical and numerical behaviors, which can be improved by increasing the measurement time.

To close this section, we would like to underline the broad feasibility of our achievements. In fact, with few simple assumptions we are able to accurately estimate the average velocity 
of a particle escaping from a metastable state of a cosine potential with friction, in the presence of a driving force and L\'evy distributed fluctuations.

\section{Finite temperature effects}
\label{Sec04}\vskip-0.2cm

\begin{figure}[t!!]
\centering
\includegraphics[width=\columnwidth]{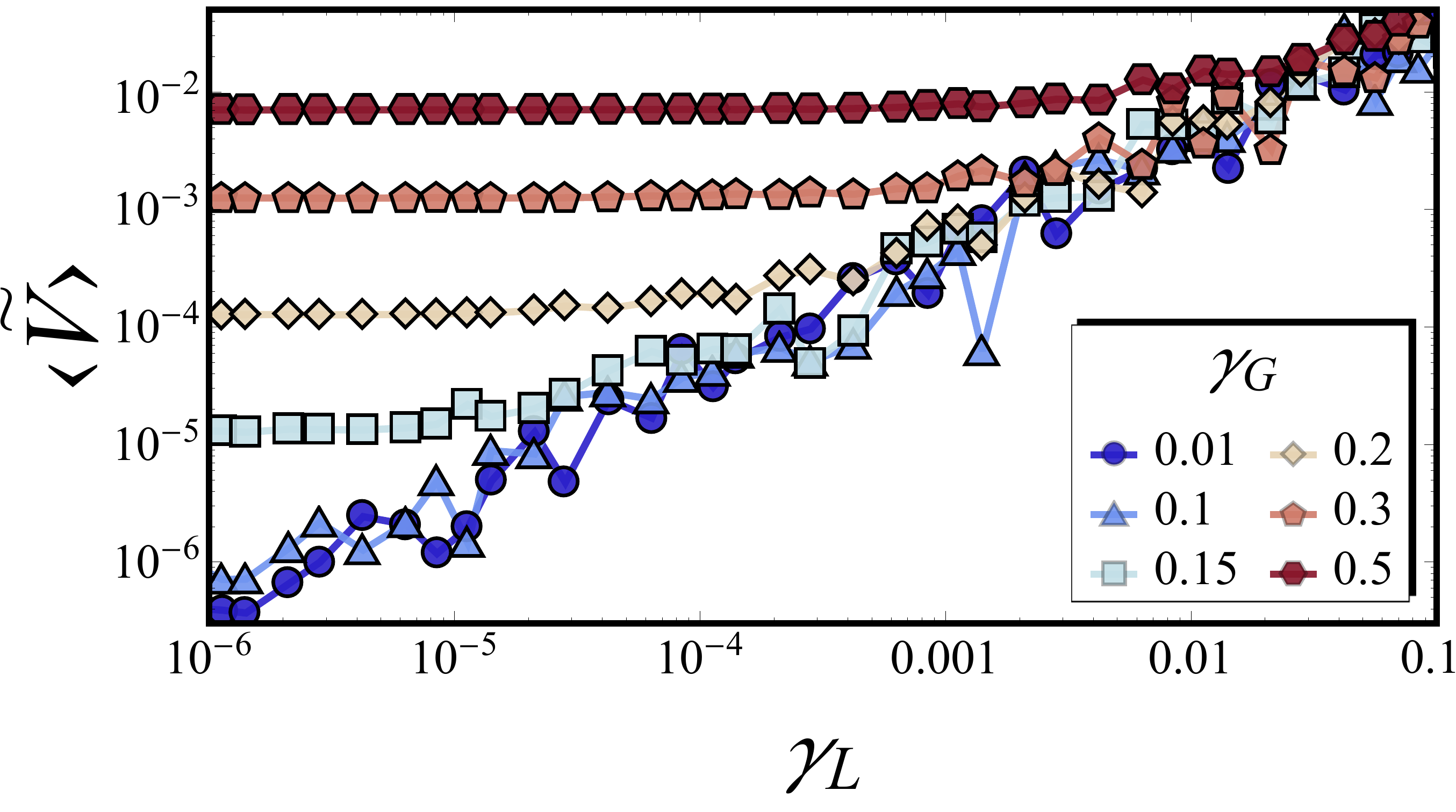}
\caption{Normalized average voltage drop as a function of the amplitude $\gL$ of the L\'evy noise source in the short-junction case, at $\alpha=1$ and $i_b=0.2$, in the presence 
of a Gaussian noise source with amplitudes $\gG=\{0.01,0.1,0.15,0.2,0.3,0.5\}$. The lines in the figure are guides for the eye.}
\label{Fig05}
\end{figure}

In this section we demonstrate that our detection method remains quite compelling also if the L\'evy component is embedded in a thermal noise background. In the proposed 
scheme the temperature of the system is a disturbance, for the contemporary presence of both the L\'evy and the Gaussian noise source with a non-negligible 
amplitude ($\gG\ne0$) entails a deviation from the expected linear behavior of the voltage as a function of the L\'evy noise amplitude. The $\V$ vs $\gL$ data shown in Fig.~\ref{Fig05}, 
obtained at a fixed L\'evy noise index $\alpha=1$ and a bias current $i_b=0.2$, changing the Gaussian noise amplitude $\gG$, demonstrate how the $\V$ response depends 
on the additional Gaussian contribution. For $\gG\lesssim 0.1$, thermal noise has no effects on the average voltage drop and $\V$ follows the linear behavior already discussed 
in Fig.~\ref{Fig02}. Conversely, a $\V$ plateau, whose value increases with $\gG$, develops for thermal noise $\gG> 0.1$. In other words, at low $\gL$ values the 
phase dynamics is dominated by the Gaussian contribution and it is therefore independent of the $\gL$ value.

The deviations from the pure-L\'evy noise case at noise amplitude $\gG> 0.1$ can be estimated from Kramers rate. In fact, for a bias current $i_b=0.2$ and a measurement 
time $t_{\text{max}}=10^4$, the condition $r(i_b,\gamma_{_G}^{\text{th}})=\tau^{-1}_{\text{max}}$, where $r$ denotes the Kramers escape rate of Eq.~\eqref{Kramersescape}, gives a noise 
amplitude $\gamma_{_G}^{\text{th}}\simeq0.096$. Therefore, it is reasonable to expect that a noise amplitude $\gG\lesssim 0.1$ does not affect the voltage response within the measurement 
time $t_{\text{max}}$. In this case, the main contribution arises from the L\'evy noise term, and the detection method proves to be robust against thermal disturbances. However, the level of 
Gaussian noise that leaves the system dominated by L\'evy noise depends on the time taken to perform the voltage measurement. In fact, within the time $t_{\text{max}}$ during which 
the voltage is measured, the JJ is exposed to thermal noise. The longer this exposure, the lower the temperature at which a significant number of thermal escapes occurs, escapes that 
disturb the switching processes induced by L\'evy noise that we wish to characterize.

These ideas together with the Kramers prediction allow, for a given measurement time $t_{\text{max}}$ and bias current $i_b$, to estimate the threshold Gaussian 
noise amplitude, $\gamma_{_G}^{\text{th}}$, which has no effects on the detection procedure. This estimation of the threshold value $\gamma_{_G}^{\text{th}}$ is possible also for a range of 
measurement times which is prohibitive for numerical simulations. In detail, through Eq.~\eqref{WNAmp} one can also evaluate the maximum working temperature for an effective detection. 
This limit can be defined as the highest temperature that does not affect the voltage, that is the temperature at which the Gaussian noise amplitude implies that the inverse Kramers rate 
equals the measurement time. 
%

\begin{figure}[t!!]
\centering
\includegraphics[width=\columnwidth]{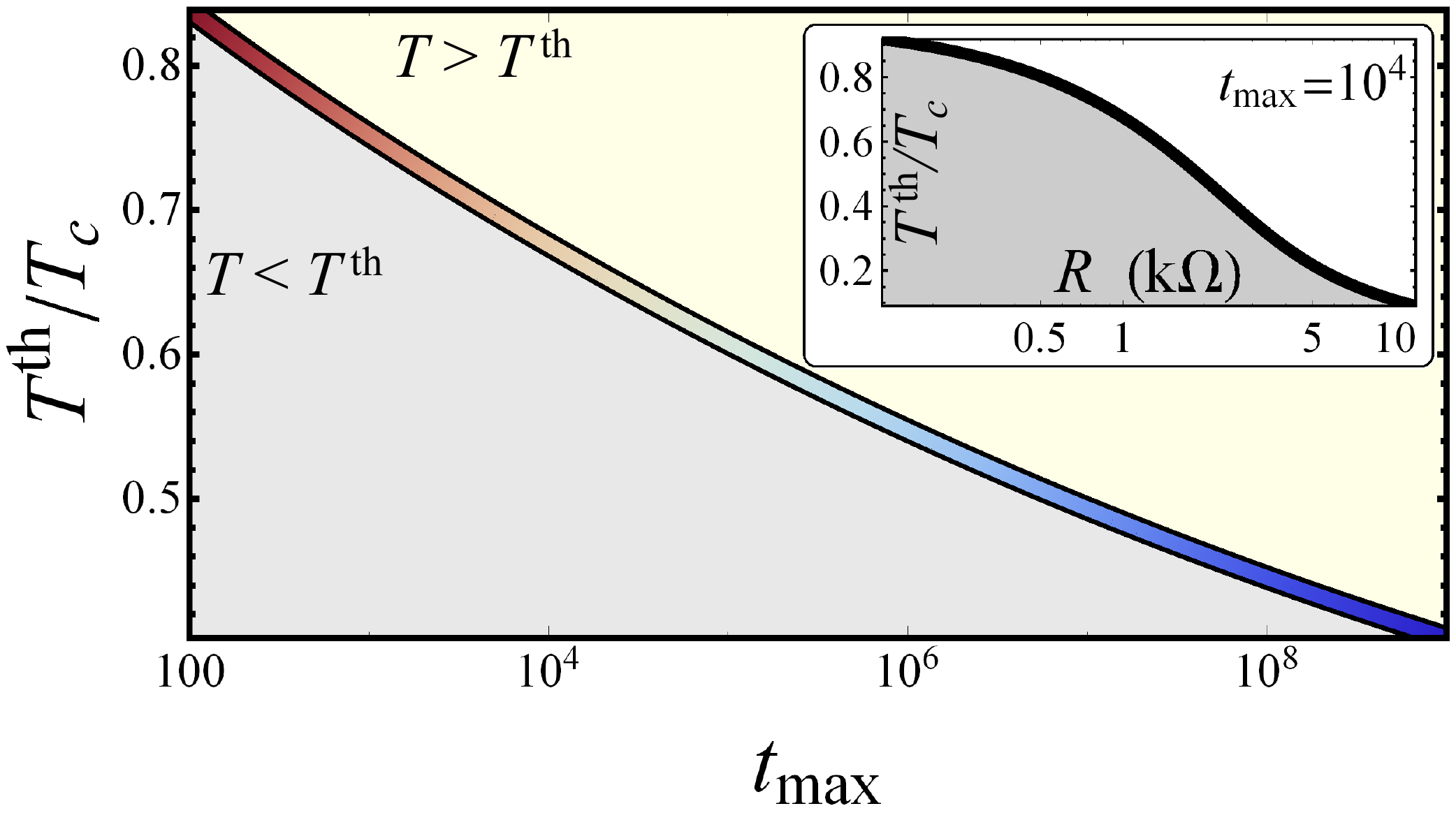}
\caption{Normalized threshold temperature, $T^{\text{th}}/T_c$, as a function of the normalized measurement time, $t_{\text{max}}$. The normalized bias current is $i_b(0)=0.2$, 
and the normal-state resistance $R=1\;\text{k}\Omega$. The inset displays $T^{\text{th}}/T_c$ versus $R$ at a fixed $t_{\text{max}}=10^4$.}
\label{Fig06}
\end{figure}

To compute this threshold working temperature, $T^{\text{th}}$, one should take into account a temperature-dependent critical current $I_c(T)$, for instance following the well-known 
Ambegaokar-Baratoff relation~\cite{Amb63}. At a fixed physical bias current $I_b$, the normalization deserves some attention, inasmuch the critical current, and therefore also the normalized bias current, depends on the temperature, \emph{i.e.}, $i_b(T)=I_b/I_c(T)=i_b(0) I_c(0)/I_c(T)$. The estimated threshold temperature $T^{\text{th}}$, in units of critical temperature 
$T_c$, as a function of the measurement time $t_{\text{max}}$ is shown in Fig.~\ref{Fig06}. Here, we have chosen the values of the low temperature bias current, \emph{i.e.}, 
$i_b(0)=0.2$, and the normal-state resistance $R=1\;\text{k}\Omega$. In this plot the gray shaded region denotes the temperature range $T<T^{\text{th}}$ where the detector 
can work ``safely'', \emph{i.e.}, without significant thermal disturbances. Instead, the yellow shaded region in Fig.~\ref{Fig06} for $T>T^{\text{th}}$ indicates the parameter region for 
which thermally-induced changes in the voltage response could hinder the accurate estimation of the characteristics of the L\'evy component.

To give figures, if the voltage measurement is performed in a normalized time $t_{\text{max}}=10^9$ (in physical units, this is a time of the order of milliseconds if $\omega_c\sim1\;\text{THz}$), 
according to Fig.~\ref{Fig06} the working temperature can be set to values $T\lesssim0.2\;T_c$ with negligible temperature-induced disturbances on the detection.

The range of suitable temperatures can be also adjusted assuming a junction with a different normal-state resistance $R$, that also affects the critical current which in turn 
determines the height of the potential barrier $\Delta U$. The inset of Fig.~\ref{Fig06} illustrates the behavior of $T^{\text{th}}/T_c$ as a function of $R$ at a fixed $t_{\text{max}}=10^4$ 
and $i_b(0)=0.2$. It is evident that the working temperature reduces monotonically with a larger normal-state resistance of the junction. 

%

\section{Conclusions}
\label{Sec05}\vskip-0.2cm

We propose to characterize the features of a L\'evy noise conveyed to a Josephson junction. We have shown that in these circumstances the average voltage 
drop across a short tunnel JJ is sensitive to the presence of such a L\'evy noise source, characterized by a fat-tails distribution, \emph{i.e.}, by a finite probability of a fluctuation 
with infinitely large intensity. The average voltage drop exhibits a peculiar behavior as a function of the noise amplitude, which is markedly different from the Gaussian noise case, 
because of the L\'evy flights, that is scale-free jumps. Specifically, the voltage grows linearly as a function of the L\'evy noise amplitude and exponentially in the Gaussian case. 
Therefore, if the noise source feeding the JJ can be attenuated, it would be possible to observe a linear behavior, markedly different from the expected response 
to a Gaussian noise. Moreover, we show that the slope of the linear behavior depends on the L\'evy index $\alpha$, and it is therefore possible to 
discriminate a feature of the noise source from the analysis of the junction voltage. The proposed method proves to be particularly effective for $\alpha\gtrsim1$, 
while remaining valid for $\alpha \lesssim 1$ and can be considered a generalization of the approach previously proposed~\cite{Gua19} based on the study of switching current 
distributions, that instead was demonstrated to be especially valuable in the region $\alpha < 1$.

To optimize the detection we have analyzed the tunable parameters. In particular, the influence of the constant bias current on the detection scheme has been examined, and we have observed that the method is most effective at a low bias current. Moreover, thermal effects can be made marginal if the device 
temperature is kept below a certain threshold. This limit temperature at which the Gaussian noise becomes negligible has been estimated, and it is in nice agreement 
with simulations. Therefore, the proposed method can be made quite robust in recognizing the L\'evy component also in a noisy, \emph{e.g.}, thermal, background, 
especially at small bias currents.

Finally, we also give an analytical expression of the average velocity, $\left \langle v \right \rangle$, of a particle in a metastable 
washboard potential under the influence of L\'evy-distributed fluctuations, with $\left \langle v \right \rangle$ corresponding to the voltage in the Josephson framework. The estimate well matches our numerical results, thus allowing for the application to overdamped 
diffusion in a tilted potential~\cite{Sokolov09,Liu19}.

A further comparison between the setup of Ref.~\cite{Gua19} and that discussed in this work is useful. 
The two methods require different setups: ac the former~\cite{Gua19} and dc that discussed in this work. In particular, in Ref.~\cite{Gua19} the JJ is assumed to be biased by an increasing current, whose profile plays a role in the construction of the switching current distributions (SCDs). Instead, in the present work we propose to bias the junction with a constant current below the critical value. Therefore, in this case an accurate and stable dc input current is required. Moreover, in Ref.~\cite{Gua19}, in order to correctly construct a SCD with a large enough number of events, each single sweep of the bias current has to be sufficiently slow. The speed at which the bias current is swept is limited by the electronics and by the occurrence of unwanted (i.e., not induced by noise) premature switching. A further complication is introduced by the ``reset'', that has to be performed to drive the bias current down to zero after each switching. Instead, the present proposal does not present such a difficulty, being based on the stationary velocity (which corresponds to a steady average voltage value) of a phase particle moving in a tilted washboard potential. In this case, the tilting has to be stable, so it calls for the different requirement that the dc current is steady enough. 
Finally, the proposal~\cite{Gua19} needs the construction of a SCD through many independent repetitions. Instead, the present proposal works in practice with a single, long enough stochastic time series; indeed, performing many independent numerical repetitions is a computational expedient that allows also for an extensive parallelization. In a concrete experiment, this could be not the case and it may suffice to study a single noise signal.

By way of conclusion, it is conceivable that the analysis of the voltage response of a JJ paves the way to the concrete application of Josephson devices 
for characterizing L\'evy noise sources. We speculate that the issue of concrete experimental estimates of the characteristic L\'evy parameters is a further, not yet fully explored, 
extension of the potentialities of Josephson-based noise detectors.

\begin{acknowledgments}
\indent This work was supported by the Government of the Russian Federation through Agreement No. 074-02-2018-330 (2), and partially by Italian Ministry of University 
and Research (MIUR).
\end{acknowledgments}


%

\end{document}